**Intrinsic motivation, Need for cognition, Grit, Growth Mindset and Academic**

**Achievement in High School Students: Latent Profiles and Its Predictive Effects**

Jun Wu, Shuoli Qi* and Yueshan Zhong*

Shanghai Institute of Early Childhood Education, Shanghai Normal University, Shanghai,

China

SILC Business School, Shanghai University, Shanghai, China

Department of Biological Science, Xi'an Jiaotong-Liverpool University, Suzhou, China

*These authors contributed equally to this work.

**Author Note**

Correspondence concerning this article should be addressed to Nancy Tsai,

McGovern Institute for Brain Research, Massachusetts Institute of Technology, Cambridge, MA 02139, United States. Email: ntsai@mit.edu

## Abstract

Recent efforts to identify non-cognitive predictors of academic achievement have especially focused on self-constructs, whose measurement is concerned with a specific domain (e.g., mathematics). However, other important factors, such as character and motivation, have received less attention. Additionally, the predictive accuracy of non-cognitive factors lacks evidence from subjects including English and Science. In this study, we take a person-centered approach and focus on students' intrinsic motivation, need for cognition, grit, and growth mindset. We mainly focus on how these factors predict students' mathematics, English, and science grades between 9th grade and 12th grade. 2,308 samples from high school students in Boston (Female = 1,237; aged from 13 to 17). The research results indicated that: (1) four latent profiles of students emerged: High in grit students ($n$ = 997, 43.2%, higher scores of grit); Moderate students ($n$ = 905, 38.3%, moderate in all scores); High in intrinsic motivation students ($n$ = 252, 11.8%, higher scores of intrinsic motivation); Low in grit students ($n$ = 154, 6.7%, lower scores of grit); (2) students' gender, race, maternal education level, and social-economic ranking predicted the profiles; and (3) four profiles of students had a significant predictive effect on Mathematics, Science and English scores in both 9th grade and 12th grade. We discussed the importance of character education for adolescents and motivation for learning in high school.

**Intrinsic motivation, Need for cognition, Grit, Growth Mindset and Academic**

**Achievement in High School Students: Latent Profiles and Its Predictive Effects**

Education is one of the important elements for improving the quality of human life in the 21st century, and academic achievement gains much attention in the public debate. There are many factors that could potentially affect students' performances. For improving academic achievement, non-cognitive variables have been taken into consideration in the previous study (Lazar et al., 2014), self-efficacy, anxiety, confidence, and self-concept.

Although there was a lot of research that took efforts in exploring how cognitive and non-cognitive factors related to student's academic outcomes, there is still much to be considered to iterate approaches. In this research, we tried to use latent profile analysis (LPA) to figure out motivational patterns within individuals and how they affect academic outcomes. Compared with traditional methods (such as cluster analysis and multiple regression), LPA is a more personal-centered approach while traditional techniques analyze data in a variable-centered way (Dena et al., 2006). Huang, Wang, & Hsu (2011) pointed out that this method uses continuous variables that represent individuals on a capacity scale, and each indicator is independent of any other latent categorical variables. In this case, LPA could benefit us to figure out the effects of different levels of variables, which opposes to a purely linear effect in the traditional way.

Moreover, more characteristics were found to be possible indicators for predicting the actual proficiency of students, including intrinsic motivation, need for cognition, grit, and growth mindset. These four non-cognitive factors play an important role in student's academic achievements. Additionally, several empirical studies investigated correlations between the performance of students and their efforts (Anders et al., 2021), but few studies went further to explore how these four potent non-cognitive predictors (intrinsic motivation, need for cognition, grit, and growth mindset) make a difference among students in a more person-centered approach.

**Intrinsic motivation**

In education research, researchers and educators are interested in the motivational patterns that students hold regarding their process of learning, and intrinsic motivation is an important element of motivational patterns that drives behaviors by internal rewards. Intrinsic motivation also indicates the impacts that arise from a person that inspires him (or her) to behave (Bomia, 1997). A higher level of intrinsic motivation could help people to be more willing to perform the work with effort while staying efficaciously (VandeWalle, 2001), and the spontaneous tendency to seek out challenges and novelty, to explore and exercise one's capacity would potentially benefit students' academic outcomes. A high rate of intrinsic motivation could contribute to oneself being more willing to fulfill the work with efforts, to interact in intention putting tactics, and meticulously plan the work. Additionally, intrinsic motivation could also be beneficial for individuals to stay steadfast after one encounters venture-associated difficulties (VandeWalle, 2001). A person who has a high intrinsic motivation is more possible to exert great attempt of her/his work, then meticulously plan their work. Because of those traits, intrinsic motivation could effectively predict the performance of individuals (Janssen and Van Yperen, 2004).

**Need for Cognition**

Cognition is related to mental processes including comprehension, gaining knowledge, and higher-level functions of the brain and previous studies defined the Need for Cognition as an individual's tendency to interact in and experience effortful cognitive endeavors (Cacioppo, Petty, Feinstein & Jarvis, 1996). People who have a high need for cognition tend to be more active processors. Over the years, there was an accumulation of studies indicating that the Need for Cognition is associated with overall performance in some obligations which might apply to cognitive processing. Since the Need for Cognition refers to one's tendency to engage in difficult cognitive activity, it could potentially contribute to students' academic outcomes. The people with a high Need for Cognition tend to take action and develop attitudes on thoughtful evaluation of information. Kreitner & Kinichi (2003) pointed out that the Need for Cognition is a feature of a person to reflect on consideration of the way to use and layout movements in a good way to achieve the favored goals.

**Grit**

Grit is another factor considered by the researchers in terms of academic performance. It was defined as passion or persistence for one's goals (Hill, Burrow, & Bron, 2016), and individuals with higher grit have been shown to reach higher grades averages, higher educational levels, and greater success in school competitions (Duckworth et al. 2007; Duckworth and Quinn 2009). It is a personality trait that is related to perseverance and passion in service of a long-term target despite being confronted by distractions and obstacles. Grit is an important predictive indicator of achievement, independent of talent and intelligence contribution. Collectively, some findings endorse the success of difficult targets includes not only talents but also the sustained and targeted utility of expertise over time (Duckworth et al., 2007).

**Growth mindset**

A growth mindset performs an essential function in students' motivation for learning, which refers to people's beliefs about their characteristics, including intelligence, abilities, talent, and personality. This specific type of belief helps people realize that their skills and intelligence could be improved with their efforts and persistence, and these believes may influence their performance and motivation (Blackwell et al., 2007; Dweck, 2006). Growth-oriented mindsets play a critical role in motivating students to learn, and students who believe that their capabilities are malleable are much more likely to persist whilst encountering difficulties (Vongkulluksn, Matewos, & Sinatra, 2021). Having a growth mindset helps students to be well performed in school, and this feature could potentially improve their economic situation in the future (Card, 2001). Compared to students with fixed mindsets, for instance, students with growth mindsets are more likely to perform well in school, taking on more challenging tasks and earning better grades (Aronson, Fried, & Good, 2002; Blackwell, Trzesniewski, & Dweck, 2007; Romero et al., 2014). Additionally, parents and teachers could change children's mindsets. For example, praising children for their effort ("You worked so hard to learn this!") causes them to have growth mindsets while praising them for their intelligence ("You're so smart!") causes them to have fixed mindsets (Mueller & Dweck, 1998). Similarly, parents who see failure as helpful tend to have children with growth mindsets, whereas parents who see failure as harmful tend to have children with fixed mindsets (Haimovitz & Dweck, 2016).

**Purpose of the Present Study**

In the present study we take a person-centered approach and focus on students' intrinsic motivation, need for cognition, grit, and growth mindset. we raised research questions on how these four non-cognitive factors are associated with students' academic outcomes:

1. How many student profiles can be identified according to intrinsic motivation, need for cognition, grit, and growth mindset of 9th-grade students?

2. What are the significant predictors of students' profiles?

3. What is the relationship between different students' profiles and academic achievement? Can students' profiles in 9th grade predict their academic achievement in 12th grade?

For the first research question, we primarily suppose that distinctive profiles which indicates high/moderate/low capability in certain characteristics should be detected and yield different categories of different cognitions (for example, high in grit); in the next stage, we assume some non-cognitive factors including students' gender, race, maternal education level, and social-economic ranking could potentially devote to distinguish these latent profiles; for the last question, it is presumed that students' profiles in 9th grade could be an effective predictor to foresee their academic achievement in 12th grade, and personality & motivation profiles are expected to be pivotal predictors of 9th grade achievement for students' performance in certain subjects.

## Method

### Sample

A cross-sectional dataset was collected in this research through students from 9th-grade to 12th-grade, and a group of 2, 308 high school students in public schools took part in this research, 46.4% of them are male and 53.6% of them are female. Collected data consisted of self-report questionnaires on demographic information, and non-cognitive skills such as grit, growth mindset, intrinsic motivation, and need for cognition. Moreover, academic performance data was included in math, english, and science.

### Instrument

#### *The Short Grit Scale.*

This measurement assesses perseverance and orientation toward long-term goals,

and it is a short version of the GRIT-O measure, where 12 items were included on the same scale (Duckworth et. al., 2007). Additionally, four items that fell below the median in prediction were deleted from the original measure. This could increase the goodness of fit and reliability across the measure. In this study, students reported their own grit by completing a 5-point Likert scale ranging from 1 (very much like me) to 5 (not like me at all), and the overall score is calculated by averaging all answers.

***Intrinsic Motivation Inventory.***

This measurement assesses students' subjective experience regarding target

activity in experiments, which is a multidimensional measurement technique to measure

intrinsic motivation. A 9-item version with three subscales were included in this study:

pressure/tension, perceived competence, and interest/enjoyment, and the Likert scale is ranging from strongly disagree (1) to strongly agree (7), higher scores stand for higher intrinsic motivation.

***Growth Mindset Scale***

This self-reported scale measures the degree of belief of people to get smarter if they focus on their work, with a 6-point scale (1 = strongly agree; 6 = strongly disagree), and it is developed by Carol Dweck (1999, 2006). High growth mindset people always believe that they could get smarter with efforts, while fixed mindset people believe that they could hardly change their mindset.

***MacArthur Scale of Subjective Social Status- Youth Version.***

This scale measures how young person perceives their own and family's social standing consist of two item instruments. The first item of the scale was used in this study for respondents to think of their family at the ladder with 10 rungs, then mark the rung which best represents the situation of their family.

**Data Analysis**

First, to address RQ1(How many student profiles can be identified according to intrinsic motivation, need for cognition, grit, and growth mindset of 9th-grade students?), we employed the latent profile analysis (LPA) to classify the students into subgroups based on scores of intrinsic motivation, need for cognition, grit, and growth mindset variables via Mplus 8.6. LPA is a probabilistic, or model-based, variant of traditional cluster analysis (Vermunt and Magidson 2002). The goal of LPA is to group individuals into categories, each one containing individuals who are similar to each other and different from individuals in other profile classes (Muthén and Muthén 2000). We selected the best-fit model based on the following characteristics of model-fit indices:(1)a significant p-value of the Lo-Mendell-Rubin likelihood ratio test (LMR-LRT; Lo, Mendell, Rubin,2001), which was developed to distribute samples for the likelihood ratio test with better accuracy;(2)a relatively low value of Bayesian information criterion (BIC; Schwartz,1978)statistics which were computed based on the model-based log-likelihood statistics (values closer to zero means better-fit model), the number of parameters, and sample size;(3)a relatively low value of sample-size-adjusted BIC (ABIC)which can weaken the penalty factor and promote the performance of BIC; and (4)a relatively high value of entropy which could indicate the accuracy of model-based classification (Peugh Fan,2013).

Second, to address RQ2(What are the significant predictors of students' profiles?), we conducted multinomial logistic regression (MLR) analyses to examine the associations between students' profiles and their demographic backgrounds (including their gender, race, maternal education level, and social-economic ranking) using Mplus 8.6.

Third, to address RQ3(What is the relationship between different students' profiles and academic achievement? Could students' profiles of 9th-grade predict their academic achievement in 12th grade?), the multinomial logistic regression analyses were conducted using the students' academic achievement in mathematics, English, and science to examine how these particular student's profiles could predict students' academic achievement between 9th-grade and 12th-grade by Mplus 8.6. In addition, a series of variances (ANOVAs) on SPSS.26, was used to estimate differences in academic achievement between 9th grade and 12th grade students.

## Results

## Four emergent profiles of students

Table 1 presents the results of five different models based on statistical analyses of the latent profiles. When the students are classified into two profiles, the two profiles would be roughly broken down by 56.80% and 43.20% of students in each group. The Entropy is 0.993, close to1, and the pLMR and pBLRT are close to 0, which means the result is satisfied. Nevertheless, the AIC, BIC, and ABIC of model C = 2 are relatively high. When there are three profiles, the percentages of the total sample size included in each of the three groups are 6.67%, 50.13%, and 43.20%. The pLMR of the model is .054, indicating that the model is less accurate. In model five in Table 1, respectively occupying 6.67%, 36.56%, 34.10%, 13.56%, and 9.11% in each profile, the Entropy becomes lower. As a result, the fit index for latent profile analysis of the fourth model is an optimal option. The four groups would contain 39.21%, 6.67%, 43.20%, and 10.92% respectively. The Entropy is 0.917. The pLMR and pBLRT are close to 0.

The four lines in Figure 1 reflect the characteristics of different profiles regarding the growth mindset, grit, intrinsic motivation, and the need for cognition. For the first profile in the yellow line, the group is characterized by particularly low grit. The mean of grit in the latent profile is just approximately 2. The overall mean scores in the other three aspects are also low. The group is quite extreme, including 6.7% of students. High intrinsic motivation student is the second latent student profile and is represented in the green line in the figure. The statistics of intrinsic motivation is over 6, reaching a peak in the figure. Additionally, the need for cognition in the group is also high. However, high in intrinsic motivation student is average in grit. The third group is high in grit, depicted in the blue line in the figure. The mean of grit in the group reaches 4. The group covers 43.2% of the students. For the part of moderate students shown by the red line, the group is relatively at an average level in each aspect despite the low need for cognition in the group. The mean of the need for cognition in the group is nearly 0.

**Predicating students profile**

As part of research question 2, we aim to examine the differences in demographic factors between the four different groups. Setting the Moderate students as the reference group, the effects of various demographic factors on different groups are in Table 2. The demographic factors include Gender, Race, Maternal education, and Social-economic ranking.

When comparing the low grit students and moderate students, the differences in maternal education are significant (*Logit* = -.774), and the difference in race is also relatively significant (*Logit* = -.774). In terms of academic achievement, there are significant differences between the two groups in grade 9 science scores and grade 12 math scores. In the comparison of high grit students and moderate students, the gender and social ranking exist a significant difference in the two groups, while the difference between grade 9 math score, grade 9 science score, grade 12 science score, and grade 12 English score is significant. For high intrinsic motivation students and moderate students, the difference in race, social-economic ranking, and grade 9 science score is significant.

**Examining the academic achievement between profiles**

Students in various profiles tend to differ in their academic outcomes. The group of Moderate students was set as the reference group in RQ3. Multinomial logistic regression is adopted based on the LPA outcome to investigate the potential predictors of the students' profiles via a three-step approach in Mplus. Accordingly, the profiles of students have a significant predictive effect on Math and Science scores in 9th grade. The results show that the students' low need for cognition tends to score higher in science ($p$ = .000), while intrinsic motivation would have negative effects on the science score ($p$ = .001). In addition, students with low grit tend to have lower English scores ($p$ = .031).

Motivational and character profiles of 9th-grade students could predict their academic achievement in 12th grade to some extent. The High intrinsic motivation students are set as the reference group in this study. According to the outcome in Figure 2, the profiles of students have a significant predictive effect on Science and English scores in 12th grade. Students low in grit tends to be significantly correlated with English and Science performance in 12th grade ($p$ = .000). Students low in need for cognition and high in intrinsic motivation tends to correlate significantly with the English score in 12th grade.

It could be observed in Figure 2 that the students low in need for cognition, high in intrinsic motivation, and high in grit tends to decline in math score while students low in need for cognition decline much more than the other two profiles. However, the performance of students low in grit tends to increase substantially. For science score, students low in grit or high in intrinsic motivation tends to surge, while students low in need for cognition and high in grit tends to decrease. For the English score, the low grits students show greater potential and the score tends to be higher than any other profiles. For students low in need for cognition and high in intrinsic motivation, the performance becomes better in 12th grade. The only profile that presents a declining trend is students high in grit.

## Discussion

In this study, a sample of 2,308 high school students in Boston was selected to explore the profiles of character and motivation among non-cognitive factors, their potential predictors, and the predictive effects of these profiles on students' academic performance. This section will discuss these findings and their implications in a pedagogical context.

## Four distinct profiles of students

The LPA technique used in this study yielded four profiles of students' characters and motivation among adolescents, which have distinctive features. The highest proportion of students were those with high grit ($n$ = 997, 43%), who scored highest in grit, while they scored lower than average in intrinsic motivation and need for cognition, especially in the growth mindset. Thus, the "portrait" of most of the students in this study was that they had the character of perseverance but lacked the motivation to continue learning. In contrast, the number of students with low grit was the lowest ($n$ = 154, 6%), scoring the lowest on all indicator variables, especially in grit. The "portrait" of these students was that they had neither the desire to learn nor the character to persevere.  Students with high intrinsic motivation ($n$ = 252, 11%) scored highest on growth mindset, intrinsic motivation, and need for cognition, and were only in the middle on grit. The "portrait" of those students who had a strong interest in learning and were able to show some persistent efforts in learning. Moderate students ($n$ = 905, 38%) were average in terms of growth mindset and grit, and low in terms of both intrinsic motivation and need for cognition. In other words, the "portrait" of a significant number of students was that they had a degree of perseverance, but still do not have much motivation for learning. Notably, the 9th graders scored very closely on the growth mindset. In the indicator of grit, the scores of the four students' profiles had a great difference. In other words, grit became one of the important indicators for identifying different profiles in this study. We can find that strong motivation to study and a lack of grit often occur in individuals, and the reverse is also true. The four different kinds of student profiles show us the complexity of individual non-cognitive skills. This suggests that researchers and practitioners should identify the different characters and motivation problems of adolescents and conduct targeted interventions or training.

**Predictors of students' profiles**

The examination of the four students' profiles in relation to the potential contributing factors revealed four significant predictors, ranging from students' gender, race, maternal education level, and social-economic ranking. Further analysis reveals that demographic predictors of different student profiles vary widely, and these findings will help us better understand the factors behind different student profiles. Our findings indicate that race and maternal education level are closely associated with students low in grit. This supported that parent education plays a key role in influencing their children's grit (Trevino & DeFreitas, 2014). Parents who have lower levels of education have been found to limit their aspirations for their children and are more likely to limit discussion of aspirations and goals (Spera et al., 2009). Parents with less education may not have socialized their students toward positive grit beliefs and behaviors, leading to a lack of protection for students of less-educated parents (Katelyn R. Black, 2014). We also found that race had an effect on low grit and high intrinsic motivation students. Further statistics showed that in both groups, there were far whiter and Blacker or African students than Asian or Latino students. This may be related to the community situation of the sampled schools. Although previous studies showed that gender, race, and grit could predict first-year GPA in college (Chang, W.,2014). Girls had higher scores on grit than boys in Grades 4-8 (Rojas et al., 2012). These findings also establish a certain connection with this research, indicating that character and intrinsic motivation could be influenced by various demographic background factors. This suggests that researchers and practitioners should use a multi-angle to understand the effect of different growing environments on individual non-cognitive factors. In this study, the impact of socioeconomic ranking on student profiles is also prominent, especially in groups with high persistence and high cognitive needs. Further analysis found that students in both groups came from higher-income families. This is consistent with the findings of previous studies.

**The predictive effect of four students' profiles on academic achievement**

The results showed that character and motivation profiles were significant predictors of 9th grade Math and Science achievement. The finding supports previous research (Bowles & Gintis, 1976, Duckworth, et al., 2007, Wolter & Hussain, 2014). Mathematics achievement is regarded as an important indicator to predict students' academic achievement. In this study, the high grit group had a very significant predictive effect on Mathematics. Notably, the low in grit, high in grit, and high intrinsic motivation groups all had predictive effects on Science achievement in grade 9. In other words, non-cognitive factors of character and motivation had a greater impact on science disciplines. On the contrary, the above three groups have no predictive effect on English scores. Defined as perseverance and passion for long-term goals (Duckworth et al., 2007), grit has been recognized as an important non-cognitive factor that can lead to academic and life success. Perseverance is a characteristic that keeps individuals from giving up easily when encountering problems and challenges, while passion for long-term goals motivates them to sustain the efforts needed to achieve a goal. Individuals higher in grit have the advantage of stamina; they stay on their course despite challenges and failures, which may eventually lead them to surpass even their gifted peers who are lower in grit (Duckworth et al., 2007; Farrington et al 2012). However, this concept was challenged in this study. In examining the predictive effect of three student profiles on their 12th grade academic performance, we found that the low grit students improved significantly in 12th grade on Math and English. We assumed that the low grit students were more likely to be from disadvantaged families, with the lowest mother educational levels and the lowest socioeconomic levels. These factors affected their academic foundation when they entered high school, and their grade 9th scores in Math, English, and Science were the lowest of all student profiles. Why did the low grit students improve so much? We thought that maybe these students were influenced by their peers, motivated to learn, and gained the ability to improve their perseverance through practice. This proves that individual students' attitudes and behaviors are malleable and can prompt them to push their limits and achieve at a higher level within educational contexts (Duckworth, Peterson, Matthews, & Kelly, 2007; Farrington et al., 2012).

**Conclusion**

Four distinct profiles were presented in the profile analysis, including high grit, low grit, high intrinsic motivation, and moderate students. These profiles were related to students' gender, race, maternal education level, and social-economic ranking. We found that certain races and low maternal education levels could be devoted to low grit for students. The character and motivation profiles efficaciously predict the 9th grade Math and Science achievement. We conclude that non-cognitive factors have greater influences on science disciplines than other subjects. Additionally, students with low grit were likely to have a significant improvement by their efforts due to the effects of their surroundings.

**Limitations and Prospects**

In this study, potential profiles of students' character and motivation in grade 9 were used to test their predictive effect on student's academic performance in grade 12, without including possible changes in non-cognitive factors in grade 12. We found that some groups of students, such as high grit, high intrinsic motivation, and moderate group, had stable predictive effects. But students from disadvantaged backgrounds, the low grit group, saw a dramatic improvement in their academic performance by the 12th grade. Why did this happen? Future studies can further track this group of students and try to reveal how their grit and motivation change over time, and what the internal laws of the change are the possible influencing factors. In addition, this study found that the student profiles had a significant predictive effect on high school student's math and science performance. Future research may explore the predictive effects of non-cognitive factors such as character and motivation from the perspective of STEAM education. Participants in this study were recruited from the Boston area, and future studies will need to sample participants from a more diverse regional sample, including rural areas, as well as data from other countries.

**Table 1**

*Model fit indices for the latent profile analyses (n = 2,308)*

| Model | Npar | Log(L) | AIC | BIC | ABIC | *p*LMR | *p*BLRT | Entropy | Percentage in Profiles% |
|-------|------|--------|-----|-----|------|--------|---------|---------|-------------------------|
| C=1 | 8 | -13,299.6 | 26,615.1 | 26,661.5 | 26,635.7 | — | — | — | 100.00 |
| C=2 | 13 | -12,942.9 | 25,911.7 | 25,986.4 | 25,945.1 | 0.000 | 0.000 | 0.993 | 56.80 / 43.20 |
| C=3 | 18 | -11,743.0 | 23,522.0 | 23,625.4 | 23,568.2 | 0.054 | 0.000 | 1 | 6.67 / 50.13 / 43.20 |
| **C=4** | **23** | **-11,642.4** | **23,330.9** | **23,463.0** | **23,390.0** | **0.000** | **0.000** | **0.917** | **39.21/ 6.67 / 43.20 / 10.92** |
| C=5 | 28 | -11,589.4 | 23,234.8 | 23,395.6 | 23,306.7 | 0.0003 | 0.000 | 0.851 | 6.67 / 36.56 / 34.10 13.56 / 9.11 |

*Note.* Npar=number of free parameters; Log(L) = log-likelihood; AIC=Akaike information criterion; BIC = Bayesian information criterion; ABIC = sample size-adjusted Bayesian information criterion; pLMR = the p-value for the Lo-Mendell-Rubin adjusted likelihood ratio test; pBLRT = the p-value for the bootstrapped likelihood ratio test.

**Table 2**

*Multinomial logistic regression results of four students' profiles (n = 2,308)*

| Variables | Low in grit students VS. Moderate students | | | | High in grit students VS. Moderate students | | | | High in intrinsic motivation students VS. Moderate students | | | |
|---|---|---|---|---|---|---|---|---|---|---|---|---|
| | *Logit* | *S.E.* | *O.R.* | *95% CI O.R.* | *Logit* | *S.E.* | *O.R.* | *95% CI O.R.* | *Logit* | *S.E.* | *O.R.* | *95% CI O.R.* |
| **Demographic factors** | | | | | | | | | | | | |
| Gender | 0.091 | 0.247 | 1.065 | [-0.316, 0.497] | -0.701** | 0.208 | 0.801 | [-1.043,-0.359] | 0.370 | 0.313 | 1.170 | [-0.146, 0.885] |
| Race | -0.481* | 0.235 | 0.876 | [-0.868,-0.094] | 0.218 | 0.283 | 1.028 | [-0.247, 0.683] | -0.708** | 0.247 | 0.888 | [-1.114,-0.303] |
| Maternal education | -0.774*** | 0.188 | 0.722 | [-1.083,-0.465] | -0.378 | 0.287 | 0.930 | [-0.849, 0.094] | -0.303 | 0.358 | 0.925 | [-0.892, 0.286] |
| Social economic ranking | -0.172 | 0.254 | 0.966 | [-0.589, 0.246] | 0.677** | 0.235 | 1.064 | [ 0.291, 1.064] | 0.621* | 0.289 | 1.080 | [ 0.146, 1.097] |
| **Academic achievement** | | | | | | | | | | | | |
| Grade 9 Math score | 0.230 | 0.487 | 1.084 | [-0.571, 1.031] | -0.747*** | 0.209 | 0.804 | [-1.091,-0.403] | 0.512 | 0.386 | 1.173 | [-0.123, 1.147] |
| Grade 9 Science score | -0.879*** | 0.243 | 0.915 | [-1.279,-0.480] | 0.619* | 0.248 | 1.054 | [ 0.211, 1.028] | 0.737* | 0.306 | 1.069 | [ 0.234, 1.241] |
| Grade 9 English score | 0.356 | 0.463 | 1.023 | [-0.406, 1.118] | 0.170 | 0.311 | 1.009 | [-0.343, 0.682] | 0.516 | 0.389 | 1.030 | [-0.125, 1.156] |
| Grade 12 Math score | -0.988*** | 0.040 | 0.690 | [-1.054,-0.921] | -0.248 | 0.540 | 0.975 | [-1.137, 0.641] | -0.354 | 0.773 | 0.962 | [-1.625, 0.918] |
| Grade 12 Science score | -0.148 | 0.332 | 0.983 | [-0.693, 0.398] | 1.120** | 0.406 | 1.035 | [ 0.452, 1.788] | 1.137 | 0.587 | 1.039 | [ 0.171, 2.102] |
| Grade 12 English score | 0.218 | 0.330 | 1.025 | [-0.325, 0.761] | -1.207*** | 0.310 | 0.964 | [-1.716,-0.698] | -0.397 | 1.046 | 0.987 | [-2.118, 1.324] |

*Note.* Reference group is Moderate students, *n* = 905, 38.3%; Low in grit students, *n* = 154, 11.8%; High in grit students, *n* = 997, 43.2%; High in intrinsic motivation students, *n* = 252, 11.8%. *S.E.* = standard error; *CI* = confidence interval. *$p$ < 0.05; **$p$ < 0.01; ***$p$ < 0.001.

**Figure 1**

*Graph comparing the indicator variables score in the four-profiles solution*

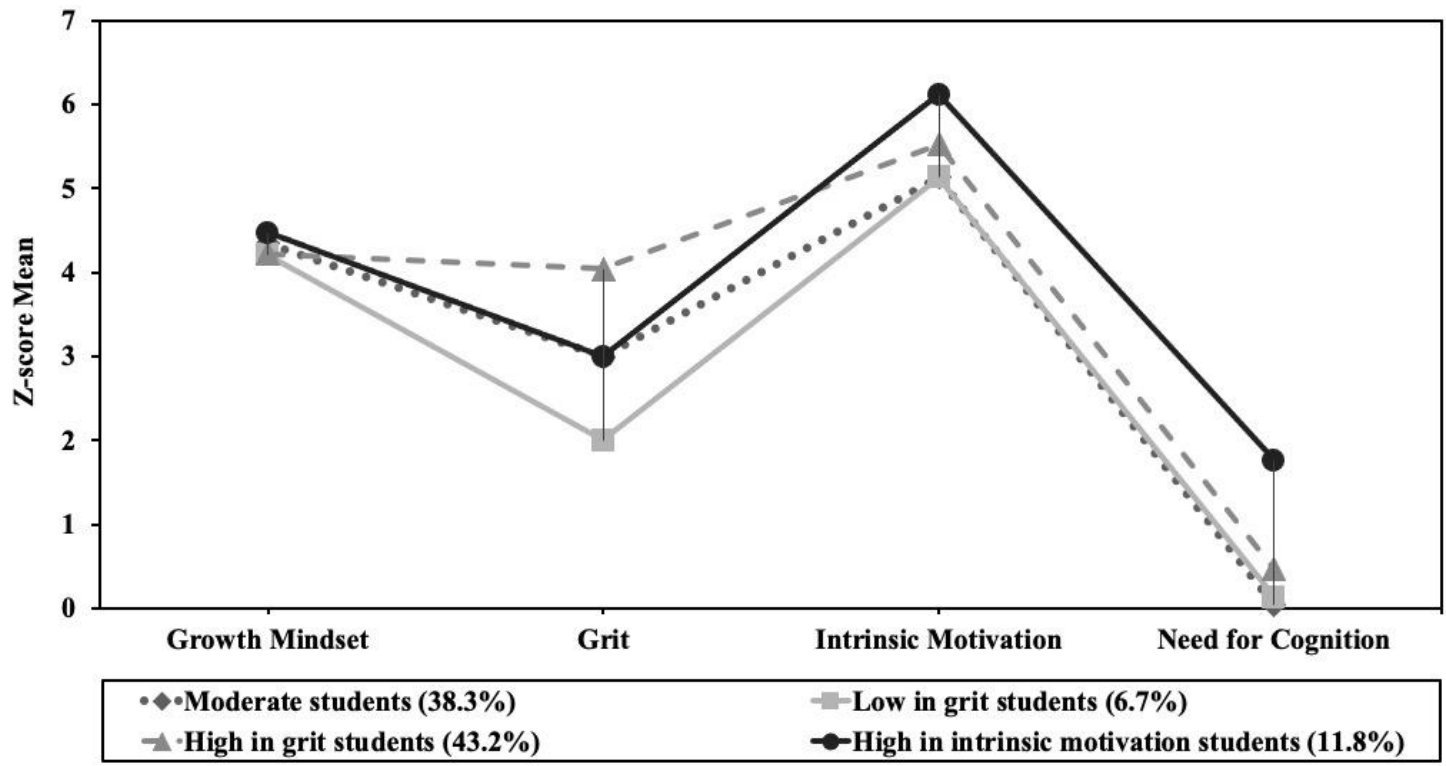

**Figure 2**

*Graph comparing the Mathematics, English and Science scores in the four-profile solution between 9th-grade and 12th-grade*

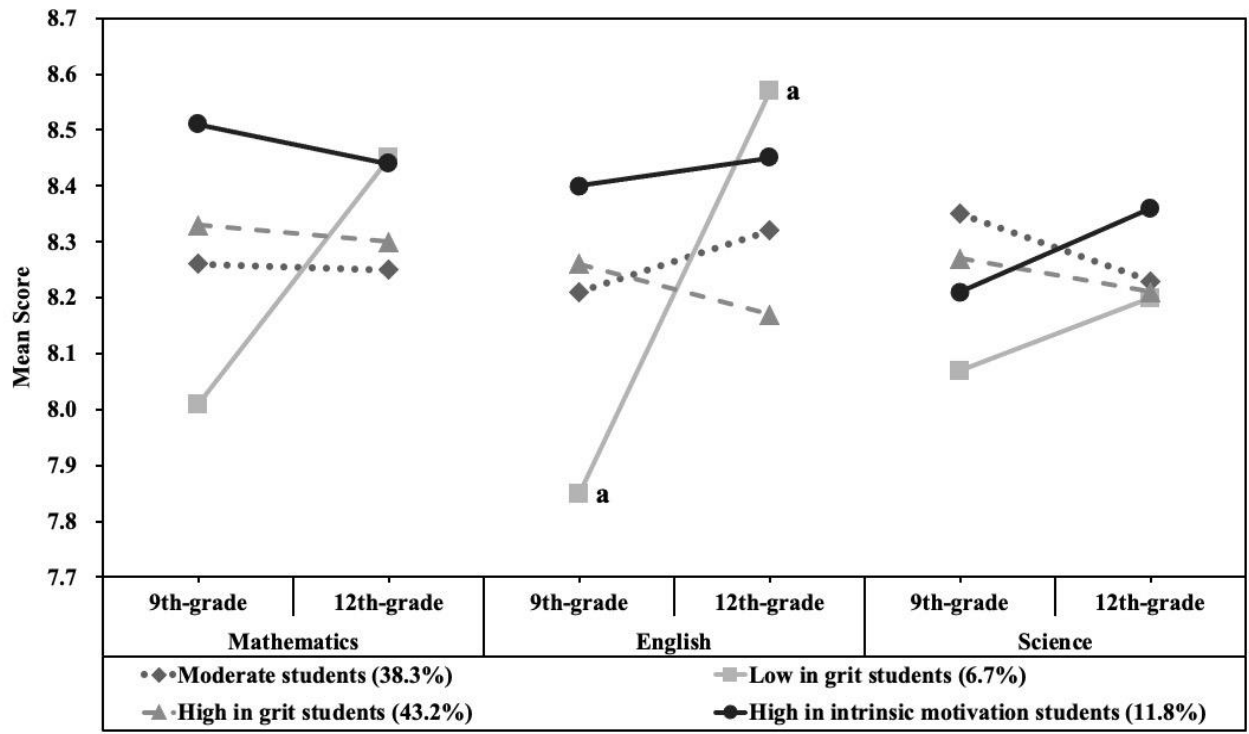

*Note.* Values with the same superscript are significantly different at α = 0.05.